\documentstyle[12pt]{article} 
\newcommand{\be}{\begin{equation}}
\newcommand{\ee}{\end{equation}}
\newcommand{\beq}{\begin{eqnarray}}
\newcommand{\eeq}{\end{eqnarray}}
\newcommand{\bear}{\begin{array}}
\newcommand{\ear}{\end{array}}

\begin{document}
\title{Schroedinger Wheeler-DeWitt Equation \\
In Multidimensional cosmology}
\author{S. Biswas,
A. Shaw and D. Biswas}
\date{}
\maketitle
\begin{center}
 Department of Physics\\
 University of Kalyani\\
 P.O.- Kalyani, Dst.- Nadia\\
 West Bengal (India)\\
 Pin. - 741235
\end{center}
\begin{abstract}
We study multidimensional cosmology to obtain the wavefunction of the universe 
using wormhole dominance proposal. Using a prescription for time we obtain the 
Schroedinger-Wheeler-DeWitt equation without any reference to WD equation and 
WKB ansatz for WD wavefunction. It is found that the Hartle-Hawking or 
wormhole-dominated boundary conditions serve as a seed for inflation as well as for 
Gaussian type ansatz to Schroedinger-Wheeler-DeWitt equation.
\end{abstract}
\newpage
\section{\bf{Introduction}}
A multidimensional spacetime seeks to explain the origin of fundamental constants 
like charge, Planck constant and Gravitational constant. As the internal 
dimensions in D+4 dimensional world are of the order of $10^{-33}\,cm$ and left 
no trace from nucleosynthesis onward, its presence obviously has to be reckoned in 
the early universe. There is now a broad consensus to introduce time in the 
framework of quantum cosmology to understand the quantum to classical transition 
of the universe and decoherence mechanism, at least to give a reasonable 
interpretation to the quantum wavefunction of the universe. The main hurdle in 
the way to interpret the wavefunction of the universe is the timeless character 
of the Wheeler-DeWitt equation that arises because of classical Hamiltonian 
constraint. Allowing a time parameter prescription, one obtains from the 
Wheeler-DeWitt equation, another equation in Schroedinger form that describes 
the evolution of the universe with respect to time. This equation is named as 
Schroedinger Wheeler-DeWitt equation. In solving the Wheeler-DeWitt equation or 
the Schroedinger Wheeler-DeWitt equation, the boundary conditions for the 
wavefunction are now taken as proposals. For Wheeler-DeWitt equation the 
proposals are namely (i). Hartle-Hawking-No boundary (HHN) proposal [1] (ii). 
Vilenkin's Tunneling proposal [2] and (iii). Wormhole-Dominance proposal [3]. 
The latter proposal has been recently proposed by us and has the features that 
in it the wavefunction is normalized, the probabilistic interpretation of 
quantum theory remains workable. Moreover both the no boundary and tunneling 
proposals can be obtained from the wormhole dominance proposal if the respective 
boundary conditions of the Hartle-Hawking and the tunneling proposals are 
introduced in it.
\par 
The wormhole dominance proposal rests on the idea of considering repeated 
reflections between the turning points present in the Wheeler-DeWitt equations. 
The normalization constant thus obtained is shown to be equivalent to wormhole 
contributions in the sense of Coleman [4] and Klebanov [5] arguments. In this 
picture `wormholes' find an interpretation as a driving agent for quantum force 
in early universe. Actually universal validity of quantum superposition 
principle requires to incorporate multiple reflections between the turning points 
irrespective of wormhole picture. The wormhole picture suggests one of the 
possibility to interpret the ``quantum force'' in the early universe.  
\par
The validity of superposition principle necessitates to explain the emergence
of classical universe from quantum state. Here we have to enforce the concept of 
decoherence i.e., the suppression mechanism of quantum interference. Though this 
concept emerged with the advent of quantum theory, in gravity the problem gets 
deeper due to absence of time in Wheeler-DeWitt equation. In Schroedinger 
Wheeler-DeWitt equation with Schroedinger like time the initial conditions are 
chosen as a Gaussian ansatz for the Schroedinger Wheeler-DeWitt wavefunction. It 
has been guessed that perhaps Gaussian ansatz is related to the boundary 
condition proposal somehow [6]. The Gaussian ansatz for the large scale factor 
region has the advantage to understand the quantum to classical transitions of 
the universe.
\par                                                                              
The motivation of the present work is to study Schroedinger Wheeler-DeWitt 
equation in multidimensional cosmology with Gaussian ansatz and investigate 
which of the boundary conditions proposal of the timeless WD equation remains 
effective with the ansatz. 
We have discussed elsewhere [7] this aspect in 4-dimensional gravity. 
\par
We start from an action in (D+4) dimensional space [8]
\be
S=\int {d^4x}\sqrt{-g}\left[-{1\over {16\pi G_N}}R+{1\over 2}
g^{\mu\nu}\partial_{\mu}\tilde{\sigma}\partial_{\nu}\tilde{\sigma}
+{1\over 2} g^{\mu\nu}\partial_{\mu}\tilde{\phi}\partial_{\nu}\tilde{\phi}
-U(\tilde{\sigma},\tilde{\phi})\right]\,.
\ee
In (1), $R$ is the four-dimensional scalar curvature, $\tilde{\sigma}$ is the dilation 
and $\tilde{\phi}$ is the inflaton and 
\be
U(\tilde{\sigma}, \tilde{\phi})=V_o(\tilde{\sigma})+e^{-D{{\tilde{\sigma}}\over \sigma_o}}V_1(\tilde{\phi})
\ee
is the combined super potential. The $\tilde{\sigma}$ field is related to the scale factor 
of the internal space through the relation
\be
\tilde{\sigma}=\sigma_o\ln {b\over b_o},\,\; where \;\;\; \sigma_o\equiv({{D(D+2)}\over 
{16\pi G_N}})^{1\over 2}\,.
\ee
The static ground state (at present) corresponds to $\tilde{\sigma}=0$, so that $b=b_o$
is the expected present size of the radius of the internal space. This requires
a potential $V_o(\tilde{\sigma})$ of the Casimir type. For details the reader is referred 
to [8,9,10]. Introducing the dimensionless fields
\be
\sigma=\sqrt{{4\pi G_N}\over 3}\;\tilde{\sigma}\,,\;
\phi=\sqrt{{4\pi G_N}\over 3}\;\tilde{\phi}\,, 
\ee
and taking 
\be
ds^2={{2G_N}\over {3\pi}}\left[dt^2-a^2(t)d\Omega_{3}^{2}\right]\,,
\ee
the action (1) after a single integration becomes 
\be
S=\int dt {1\over 2}\left[-a{\dot{a}}^2+a^3({\dot{\sigma}}^2+{\dot{\phi}}^2)
+a-a^3\overline{U}(\sigma,\phi)\right]\,,
\ee
where a dot denotes a time derivative. In (6)
\be
\overline{U}(\sigma,\phi)=V_D(\sigma) +e^{{-{D\sigma}\over {\sigma_{_{\sc{D}}}}}}V(\phi)
\ee
\[V_D(\sigma)=K\left[{2\over {D+2}} e^{-2{(D+2)\sigma}\over {\sigma_{_{\sc{D}}}}}
+e^{-{D\sigma}\over {\sigma_{_{\sc{D}}}}}-{{D+4}\over {D+2}} 
e^{-{(D+2)\sigma}\over {\sigma_{_{\sc{D}}}}}\right]\,,\]
\[K={{2(D-1)\sigma_{\sc{D}}^{2}}\over {(D+4)b_{o}^{2}}},\; 
\sigma_{_{\sc{D}}}=\sqrt{{D(D+2)}\over 12}\,,\]
and the potential $V(\phi)$ depends on the model. In ref.[8]
$V(\phi)=\lambda{\phi^4\over 4}$ is chosen for chaotic inflation. The 
Hamiltonian constraint corresponding to (6) that follows from time-time 
component of Einstein equation is 
\be
-{1\over 2}({{\dot{a}}^2\over a^2}+{1\over a^2})+{1\over 2}
({\dot{\phi}}^2+{\dot{\sigma}}^2)+\overline{U}(\sigma,\phi)=0\,.
\ee
With $P_a=-a\dot{a},\;P_\phi=a^3\dot{\phi},\;P_\sigma=a^3\dot{\sigma}$ as
momenta, the Hamiltonian constraint equation is 
\be
-{1\over {2a}}P_{a}^{2}+{1\over {2a^3}}(P_{\phi}^{2}+P_{\sigma}^{2})
-{1\over 2}a+a^3\overline{U}(\sigma,\phi)=0\,.
\ee
Identifying $P_i={{\partial S (a,\sigma,\phi)}\over {\partial q_i}}$ where 
$q_i=a,\sigma,\phi,$ the Einstein-Hamiltonian-Jacobi equation reads
\be
-{1\over 2}({{\partial S}\over {\partial a}})^2
+{1\over {2a^2}}({{\partial S}\over {\partial \phi}})^2
+{1\over {2a^2}}({{\partial S}\over {\partial \sigma}})^2
-{1\over 2}a^2+a^4\overline{U}(\sigma,\phi)=0\,.
\ee
With $P_i=-i{\partial \over {\partial q_i}}$, the Wheeler-DeWitt 
equation in the minisuperspace of coordinates $q_i=a,\sigma,\phi,$ is given by
\be
\left[\partial_{a}^{2}-{1\over a^2}(\partial_{\sigma}^{2}+\partial_{\phi}^{2})
-\omega(a,\sigma,\phi)\right]\Psi(a,\sigma,\phi)=0
\ee
where
\be
\omega(a,\sigma,\phi)=a^2\left[1-a^2U(\sigma,\phi)\right]
\ee
with $U(\sigma,\phi)=2\overline{U}(\sigma,\phi)$. In obtaining (11) a simple factor
ordering is assumed in the super Hamiltonian. In ref.[8], the authors obtained the 
following solutions with tunneling boundary conditions.
\par
\noindent
(i). Solution of nothing: 
$a^2<A_{*}^{2}(\sigma,\phi)={1\over {\sqrt{U(\sigma,\phi)}}}$
\beq
\Psi&=&a^{1\over 2}K_{1\over 4}({a^2\over 2})\nonumber \\
&\rightarrow & e^{-{a^2\over 2}}\;\;as\;\,a\rightarrow \,0
\eeq
(ii). Minisuperspace region of small $\vert \phi\vert$: 
$V(\phi)<<e^{{D\sigma}\over {\sigma_{_{\sc{D}}}}}V_D(\sigma)$
\par
(a) For $a^2V_D(\sigma)>>1\;,\,\sigma<<-{\sigma_{_{\sc{D}}}\over D},$
\be
\Psi\propto\exp\{\pm\sqrt{{{2K}\over {D+2}}}{a^3\over {3g_D}}
e^{-(D+2){\sigma\over \sigma_{_{\sc{D}}}}}\}
\ee
In obtaining (14), $V_D(\sigma)$ is approximated by the first term in (7) only.
\par
(b) For $a^2V_D(\sigma)>>1\;,\;\sigma>>{\sigma_{_{\sc{D}}}\over D},$
\be
\Psi\propto \exp(\pm i{\sqrt{K}\over {3g_D}}a^3e^{-D{\sigma\over {2\sigma_{_{\sc{D}}}}}})\,,
\ee
In obtaining (15), $V_D(\sigma)$ is approximated by the second term in (7). They
select the negative sign in (14) and (15) to satisfy Vilenkin's boundary condition.
The Vilenkin wavefunction in the region $a^2V_D(\sigma)>1$ reads
\be
\Psi\propto\exp(-{\{1+i(V_D(\sigma)a^2-1)^{3\over 2}\}\over {3V_D(\sigma)}})\,.
\ee
Eqn.(16) follows for the choice $p=-1$ for the factor ordering parameter. For
the choice $p=0$, the case with the present discussion, 
$\exp({-1\over {3V_D(\sigma)}})$ has to be introduced in adhoc way so that one
recovers the behaviour in (13). The inflaton potential $V(\phi)$ seems to play no
role in their discussion [8]. The conditions $\sigma<-{\sigma_{_{\sc{D}}}\over D}$ though
somehow includes asymmetry in super potential $U(\sigma,\phi)$, the solution (14)
cannot reproduce Vilenkin's boundary conditions as demanded in [8]. We will show
this in the present work.
\par
In section II, we derive the Schroedinger Wheeler-DeWitt equation for our model and obtain 
the solutions with a Gaussian ansatz necessary to effect the decoherence 
mechanism. In section III, we try to understand the boundary conditions and give 
some discussions in the perspective of wormhole dominance proposal.
\smallskip
\section{\bf{SWD Equation And Solution}}
We define a time operator by the relation 
\be
{\partial \over {\partial t}}=({{\partial H}\over {\partial P_i}}
{\partial\over {\partial q_i}}-
{{\partial H}\over {\partial q_i}}{\partial\over {\partial P_i}})\,.
\ee
In multidimensional cosmology both $\sigma$ and $\phi$ act like scalar fields and
acting as matter source. Using (9) we calculate 
${\partial\over {\partial t}}$
and demand that the time parameter is determined by the geometry only.
This requires
\be
{\partial\over {\partial t}}=-{1\over a} {{\partial S}\over {\partial a}}
{\partial\over {\partial a}}\,.
\ee
In obtaining (18) we identify 
$P_i={{\partial S}\over {\partial q_i}}$ 
where $q_i=a,\sigma,\phi$. Both (17) and (18) will be satisfied provided 
$V_D(\sigma)=0$ and $V(\phi)=0$ and coefficients of
${\partial\over {\partial \phi}}\;,
{\partial\over {\partial P_a}}\,,
{\partial\over {\partial P_\sigma}},
{\partial\over {\partial P_\phi}}$ vanish. This gives $S(a,\sigma,\phi)=S_o(a)$
i.e., $S_o$ is a function of $a$ only and
\be
({{\partial S_o}\over {\partial a}})^2=a^2\;.
\ee
Equation (19) is obtained by taking the coefficient of 
${\partial\over {\partial P_a}}$
term equal to zero. Hence (18) reads
\be
{\partial\over {\partial t}}=-{1\over a} {{\partial S_o}\over {\partial a}}
{\partial\over {\partial a}}\,.
\ee
The time operator defined here is a directional derivative along each of the classical
spacetimes which can be viewed as classical trajectories in the gravitational 
configuration space. From (20) we can write
\be
{{\partial S}\over {\partial t}}=-{1\over a} {{\partial S_o}\over {\partial a}}
{{\partial S}\over {\partial a}}\,.
\ee
where $S=S(a,\sigma,\phi)$. Further writing 
$S(a,\sigma,\phi)=S_o(a)+S_1(a,\sigma,\phi)$ and demanding $S_1<<S_o$, we write
(21) as  
\be
{{\partial S}\over {\partial t}}=-{1\over a} {{\partial S}\over {\partial a}}
{{\partial S}\over {\partial a}} +{1\over a} {{\partial S_1}\over {\partial a}}
{{\partial S}\over {\partial a}}\,.
\ee
We Substitute 
$({{\partial S}\over {\partial a}})^2$ from (10) in (22) and find 
\be
{{\partial S}\over {\partial t}}=-{1\over a^3}({{\partial S}\over {\partial \phi}})^2
-{1\over a^3}({{\partial S}\over {\partial \sigma}})^2+a-2a^3\overline{U}(\sigma,\phi)
+{1\over a} {{\partial S_1}\over {\partial a}}
{{\partial S}\over {\partial a}}\,.
\ee
Substituting 
$({{\partial S_o}\over {\partial a}})^2=a^2$ in (23) and neglecting
$({{\partial S_1}\over {\partial a}})^2$ term
we get absorbing the factor 2 in
$\overline{U}$ in (23)
\be
{{\partial S}\over {\partial t}}=-{1\over {2a^3}}({{\partial S}\over {\partial 
\phi}})^2
-{1\over {2a^3}}({{\partial S}\over {\partial \sigma}})^2-{{a^3\overline{U}
(\sigma,\phi)}\over 2}
+{1\over a} ({{\partial S_1}\over {\partial a}})^2\,.
\ee
In the large $a$ region 
$({{\partial S_1}\over {\partial a}})^2$ term can be neglected compared to other 
terms. With 
$p_t={{\partial S}\over {\partial t}},
p_i={{\partial S}\over {\partial q_i}},$
and upon quantization we get
\be
i{{\partial \Psi}\over {\partial t}}=\left[-{1\over {2a^3}} 
{\partial_\phi}^2
-{1\over {2a^3}}{\partial_\sigma}^2+{{a^3U
(\sigma,\phi)}\over 2}\right]\Psi\,.
\ee
This is our Schroedinger Wheeler-DeWitt equation. In the literature (25) is 
derived from the Wheeler-DeWitt equation but in our formalism the standard
Hilbert space of quantum theory can be employed and has a general viability.
We now solve (25) with Gaussian ansatz. Another important feature of (25) is
that the $\Psi$ in our case refers to full wavefunction of Wheeler-DeWitt 
equation.
\par
To solve (25), we first consider the case 
$\vert\sigma\vert<<{\sigma_{_{\sc{D}}}\over D}$.
In this case, 
\be
V_D(\sigma)\simeq {K\over 2}({\sigma\over \sigma_{_{\sc{D}}}})^2(D+4)
+O({\sigma\over \sigma_{_{\sc{D}}}})^3
\ee
so that 
\be
U(\sigma,\phi)\simeq \left[K(D+4)+V(\phi)D^2\right]{1\over 2}
({\sigma\over \sigma_{_{\sc{D}}}})^2-V(\phi){\sigma\over \sigma_{_{\sc{D}}}}+V(\phi)\,.
\ee
\par
In expanding $V_D(\sigma)$, the coefficient of ${\sigma\over \sigma_{_{\sc{D}}}}$
vanishes and is an important feature of the super potential. Further, since
${V''}_D(\sigma)$ around $\sigma=0$ is greater than zero, $V_D(\sigma)$ has a local
maxima around $\sigma=0$.
From (27) it is clear that we can express $U(\sigma,\phi)$ as
\be
U(\sigma,\phi)=\lambda(1+m^2\sigma^2)
\ee
in which $\lambda$ and $m^2$ are obtained from (27) with a shift of the field.
We denote the shifted field by $\sigma$ again and $V(\phi)=const.=V$.
\be
m^2={(K(D+4)+VD^2)\over {2\lambda}}\,,
\ee
\be
\lambda=V(\phi)-{V^2\over {2\left[K(D+4)+V(\phi)D^2\right]}}\,.
\ee
Equation (25) now reads
\be
i{{\partial \Psi}\over {\partial t}}=\left[-{1\over {2a^3}}\partial_{\phi}^{2}
-{1\over {2a^3}}\partial_{\sigma}^{2}+{{a^3\lambda}\over 2}(1+m^2\sigma^2)\right]
\Psi\,.
\ee
Let us choose Gaussian ansatz for the SWD Equation (31)
\be
\Psi=N(t)e^{-{{\Omega(t)}\over 2}\sigma^2}e^{\lambda_o\phi}\,.
\ee
Substituting (32) in (31), we get
\be
i{d\over {dt}}\ln {N}= {\Omega\over {a^3}}-{\lambda_{o}^{2}\over a^3}+a^3\lambda
\ee
\be
-i\dot{\Omega}=a^3\lambda m^2-{\Omega^2\over a^3}\,.
\ee
With the ansatz
\be
\Omega=-ia^3{\dot{y}\over y}
\ee
one finds from (34) for $y$ the equation
\be
\ddot{y}+3{\dot{a}\over a}y+m^2\lambda y=0\,.
\ee
This can be simplified by introducing the conformal time co-ordinate $\eta$
according to $dt=ad\eta$,
\be
y''+2{a'\over a}y'+m^2\lambda a^2y=0\,,
\ee
where prime denotes the derivative with respect to $\eta$. In conformal time
we take $a(\eta)=-{1\over {\sqrt{\lambda}\;\eta}}$ as an inflationary background. We
evaluated $\Omega$, knowing solution of $y$ as 
\be
y=\eta^{{3\over 2}-\sqrt{{9\over 4}-m^2}}
\ee
and using the fact that $m^2<<{9\over 4}$. We find
\be
\Omega=i{{m^2a^3\sqrt{\lambda}}\over 3}\,.
\ee
Using the fact that we are considering solutions for large $a$ region i.e.,
$a^2V_D(\sigma)>1$, it is justifiable to keep only $a^3\lambda$ term in (33).
As
\be
{d\over {dt}}=\sqrt{\lambda}a{d\over {da}}\,,
\ee
we get from (33)
\be
\ln{N}=-i{{a^3\sqrt{\lambda}\over 3}}-\ln N_o\,.
\ee
Hence
\beq
\psi&=&N_oe^{-i{{a^3\sqrt{\lambda}}\over 3}(1+{1\over 2}m^2\sigma^2)}\nonumber \\
&=&N_oe^{-i{{a^3}\over 3}\left[\lambda(1+m^2\sigma^2)\right]^{1\over 2}},
\;since\; \sigma<<1\nonumber \\
&=&N_oe^\{{-{i\over 3}\left[a^2\lambda(1+m^2\sigma^2)\right]^{3\over 2}
({1\over {\lambda(1+m^2\sigma^2)}})}\}\nonumber \\
&\simeq &N_oe^{-{i\over {3U}}\left[a^2U-1\right]^{3\over 2}}\nonumber \\
&= &N_oe^{-{i\over {3U}}(a^2U-1)^{3\over 2}}\;.
\eeq
To evaluate $N_o$, we turn back to our `wormhole dominance' proposal and this
serves as an initial condition. According to `wormhole dominance' proposal [3] 
$N_o$ is given by 
\be
N_o={{\exp{iS_{eff}(a_o,0)}}\over {1-e^{2iS_{eff}(a_o,0)}}}\,
\ee
In (43),
\be
S_{eff}(a_o,0)=-{1\over {3U}}(a^2U-1)^{3\over 2}\vert_{0}^{a_o}\,,
\ee
where $a=0$ and $a_o={1\over \sqrt{U}}$ are the turning points. Evaluating (44)
we find
\be
N_o={{\exp{({1\over {3U}})}}\over {1-\exp{({2\over {3U}})}}}\,.
\ee
Hence the wavefunction reads $(a^2U>1)$
\be
\Psi\propto{{\exp{({1\over {3U}})}\left[1-i(a^2U-1)^{3\over 2}\right]}
\over {1-\exp{({2\over {3U}})}}}\,.
\ee
Continuing in the region $a^2U<1$, we find
\be
\psi\propto{{\exp{({1\over {3U}})}\left[1-(1-a^2U)^{3\over 2}\right]}
\over {1-\exp{({2\over {3U}})}}}\,.
\ee
If we leave aside the denominator in (46) and (47), the wavefunctions (46) and
(47) corresponds to the Hawking's proposal. The solutions (46) and (47) when continued
to $a\rightarrow 0$ region gives
\be
\psi\rightarrow e^{+{a^2\over 2}}
\ee
as desired in no boundary proposal. Hence the solution obtained in reference [8]
is questionable. The important conclusion arrived from the previous discussion 
is that both the `wormhole dominance' and the Hartle-Hawking proposals justifiably 
reckons inflation, unlike the claim by some authors since we arrive at (47) with 
$a=-{1\over {\sqrt{\lambda}\;\eta}}$ i.e., $a(t)=e^{\sqrt{\lambda}\;t}$, showing the 
importance of $V(\phi)$ term as well as the Gaussian ansatz for $a^2V>>1$ region.
\medskip
\section{\bf{Discussion}}
Let us try to understand the analytic continuation of the solution (14) in the 
region $a^2V_D<<1$. The approximation that leads to this form would give back
\be
\Psi\propto\exp{(\pm{1\over {{\it{G}}_DV_D(\sigma)}}
\left[a^2V_D(\sigma)-1\right]^{3\over 2})}
\ee
or,
\be
\Psi\propto\exp{(\mp i{1\over {{\it{G}}_DV_D(\sigma)}}
\left[1-a^2V_D(\sigma)\right]^{3\over 2})}\,.
\ee
The solution (50) cannot give back the solution of nothing (13) when 
$a\rightarrow 0$. However the solution (15) under analytic continuation leads to
\be
\Psi\propto\exp{(\mp {1\over {3{\it{G}}_DV_D(\sigma)}}
\left[1-a^2V_D(\sigma)\right]^{3\over 2})}\,.
\ee
The solution (51) will give the Vilenkin solution provided we add a term to it. 
The solution is 
\be
\Psi\propto\exp{\left[- {1\over {3{\it{G}}_DV_D(\sigma)}}
(1-(1-a^2V_D(\sigma))^{3\over 2})\right]}\,,
\ee
such that 
\be
\psi\rightarrow e^{-{a^2\over {2{\it{G}}_D}}}
\ee
as $a\rightarrow 0$. Hence the lower sign in (15) is the correct choice in 
the region $a^2V_D(\sigma)>>1,\, \sigma>{\sigma_{_{\sc{D}}}\over D}$ for the tunneling
proposal. In tunneling proposal the origin of the term 
$\exp{(- {1\over {3{\it{G}}_DV_D(\sigma)}})}$
in (51) is not known, though it occurs with a choice $p=-1$ in the tunneling 
proposal [11]. It is therefore necessary to study decoherence mechanism in the
tunneling proposal to arrive at a final conclusion and about the nucleation of 
the universe and classical evolution with respect to time according to SWD 
equation.   
\par
Our view is that retaining only $({\sigma\over \sigma_{_{\sc{D}}}})^2$ term in the solution
(15) would thus lead to the same answer as is found in this work and decoherence 
would be effective in large $a$ region, though the behaviour $a\rightarrow 0$ 
remain unsettled. This can only be decided through the wormhole dominance 
proposal requiring a definite prescription for the normalization constant and of 
course with an interpretation. We would like to report this aspect in details, 
shortly.
\newpage
\begin{center}
{\bf{References}}
\end{center}
{\obeylines\tt\obeyspaces{ 
1. S.W.Hawking, {\it{Nucl. Phys.}}{\bf{B239}}, 257(1984)
   \qquad S.W.Hawking and D.N.Page, {\it{Nucl. Phys.}}{\bf{B264}}, 184(1986)
2. A.Vilenkin, {\it{Phys. Rev.}}{\bf{D37}}, 888(1987)
3. S.Biswas, B.Modak and D.Biswas, {\it{Phys. Rev.}}{\bf{D55}}, 4673(1996)
4. S.Coleman, {\it{Nucl. Phys.}}{\bf{B310}}, 643(1988)
5. I.Klebanov, L.Susskind and T.Banks, {\it{Nucl. Phys.}}{\bf{B317}}, 
   \qquad 665(1989)
6. H.D.Conradi and H.D.Zeh, {\it{Phys. Lett.}}{\bf{A154}}, 321(1991)
7. S.Biswas, A.Shaw and B.Modak, `Time in Quantum Gravity', 
   \quad (communicated to IJMPD) 
8. E.Carugno, M.Litterio, F.Occhionero and G.Pollifrone, 
   \qquad {\it{Phys. Rev.}}{\bf{D53}}, 6863(1996)
9. L.M.Sokolowski, {\it{Class. Quantum Grav.}}{\bf{6}}, 59(1989)
10. M.J.Duff, B.E.W.Nelsson and C.N.Pope, {\it{Phys. Rep.}}{\bf{130}}, 
    \qquad 1(1986)
11. A.Vilenkin, {\it{Phys. Rev.}}{\bf{D37}}, 888(1988), 
    \qquad T.Vachaspati and A.Vilenkin, {\it{Phys. Rev.}}{\bf{D37}}, 898(1988).}}    
\smallskip
\section{\bf{Acknowledgment}}
A. Shaw acknowledges the financial support from ICSC World Laboratory, 
LAUSSANE during the course of the work.
 
\end{document}